%% file: Driver.tex
\documentclass[11pt,psfig,epsfig]{article}
\usepackage{amssymb,latexsym}
\usepackage{amsmath}
\usepackage{graphicx}
\usepackage{float}
\usepackage{algorithm}
\usepackage{multirow}
\usepackage{siunitx}
\usepackage[noend]{algpseudocode}
\usepackage{float}
\usepackage{url}
\usepackage{epstopdf}
\newcommand{\real}{{I\kern-.26em R}}
\newcommand{\natur}{{I\kern-.26em N}}

\begin{document}
\bibliographystyle{ieeetr}
\input{header}
\input{intro}

\input{Segmen}
\input{Conclus}

\bibliography{CE}
\end{document}

%% file: Header.tex
\title{Prediction of trending topics using ANFIS and deterministic models}
\author{Ren\'e Escalante\thanks{Universidad de Alcal\'a, Departamento de F\'{\i}sica y Matem\'aticas, Madrid,
		Spain (rene.escalante@uah.es).}
	\thanks{Corresponding author.} \and Marco Odehnal \thanks{Universidad Sim\'on	Bol\'{\i}var,
            Departamento de C\'omputo Cient\'{\i}f\/ico y Estad\'{\i}stica,
         Ap. 89000, Caracas, 1080-A, Venezuela (markodehnal@gmail.com).}}
 \date{August 7, 2021}
\maketitle
\begin{abstract}

Trending topics are often the result of the spreading of information between users of social networks. These special topics can be regarded as rumors. The spreading of a rumor is often studied with the same techniques as in epidemics spreading. It is common that many datasets may not have enough measured variables, so we propose a method for studying the general behavior of the spreaders by selecting estimated variables given by the deterministic model. In order to provide a good approximation, we implemented an adaptative neuro fuzzy inference system (ANFIS). So, in our numerical experimentations, a deterministic approach using SIR and SIRS models (with delay) for two different topics is used. Thus, the authors just applied the ANFIS model for their application and the deterministic model as the preprocessing input variable.
\\
\\
Key words: Fuzzy logic; ANFIS; Rumor propagation; Epidemiology; Deterministic models; Trending topics.
\\
\\
\textbf{MSC} 92D50, 92D30, 92D25, 92D99

	
\end{abstract}

%% file: Intro.tex
\section{Introduction}

Each day, the social networks keep on growing very quickly and therefore, the way that people spread information has changed dramatically. But of course, this may have serious 
 consequences on the individuals; for example, Del Vicario \textit{et al.} proposed a mathematical model \cite{bib:delvicario}, where it is concluded that social network users use to spread only those rumors which are related to their favorite category (which could be, for example, science or conspiracy theories). This behavior is reinforced when the users make contact with specialized communities in these topics, which may frequently lead them into situations of disinformation \cite{bib:delvicario}.

The spreading of a rumor exhibits quite similar dynamics to epidemics spreading and thus, epidemics-based techniques have been used. Many authors have researched following this line of work (see, for example, \cite{bib:dk}, \cite{bib:mk}, \cite{bib:nekovee}, \cite{bib:escalante}, and \cite{bib:escalodeh}). Thus, for instance, according to \cite{bib:nekovee}, a rumor can be seen as an \textquotedblleft infection of the mind\textquotedblright.

In the compartmental models, the population is divided in three classes: the \textbf{ignorants} ($S$), who do not know the rumor, the \textbf{spreaders} ($I$), who are currently spreading the rumor and, f\/inally, the \textbf{stifflers} ($R$), who even though they know the rumor, do not spread it. These three categories correspond to their epidemiologically equivalent \textbf{Susceptible} ($S$), \textbf{Infectious} ($I$), and \textbf{Recovered} ($R$), respectively. The compartment models are one of the best known deterministic models. The $SI$, $SIR$, $SIS$ or $SIRS$ models are examples of compartment models \cite{bib:brauer} [The population is assigned to compartments with labels: $S$, $I$, or $R$, in such a way that the order of the labels usually shows the flow patterns between the compartments.]

The standard propagation rumor model was introduced in 1965 by Daley and Kendall \cite{bib:dk}. Also, the modeling of rumor propagation has been proposed by other authors such as Rapoport (1948) and Bartholomew (1967) (see \cite{bib:daleygani} and references therein), and Zanette (2001) \cite{bib:zanette}. These models are based on stochastic processes. However, in this paper we propose a different approach.

In this work, we developed a mathematical model that provides a good f\/itting for the general behavior of the spreaders, using deterministic models and an adaptative neuro fuzzy inference system (ANFIS). As far as we know, f\/itting a rumor-spreading related dataset, using both deterministic models and ANFIS the way it is shown in this paper, has not been considered before. Additionally, we propose an original technique that allows us to f\/it the spreaders, and at the same time, to obtain quantitative information about unknown variables.

This paper is organized as follows. In Section 2, we review some basic concepts. In Section 3, we provide a brief description about our ANFIS approach using deterministic models, and its advantages. In Section 4 we present preliminary numerical experimentation to illustrate the proposed strategy. Finally, in Section 5 some concluding remarks are presented.

\section{Preliminaries}

\subsection{Deterministic Models}

In this paper two deterministic models are used for different situations. The f\/irst model is the well-known $SIR$ model (differential equations-based), which was proposed by Kermack and McKendrick \cite{bib:kermackmackendrick} to study epidemic diseases such as measles and rubella \cite{bib:brauer2012}.

\par The $SIR$ model is given by the following system of ordinary differential equations:
\begin{equation}
\begin{cases} \frac{dS}{dt}=-\alpha S I, \\
\frac{dI}{dt}=\alpha S I - \beta I, \label{SIReq} \\
\frac{dR}{dt}=\beta I,\end{cases}
\end{equation}

\par where $\alpha \mbox{ y }\beta$ are called \textit{infection rate} and \textit{recovery rate}, respectively.

 In addition to the distinction between diseases for which recovery confers immunity against reinfection ($SIR$ model) and diseases for which recovered members are susceptible to reinfection, we consider the intermediate possibility of temporary immunity described by a model of $SIRS$ type, with constant inmunity (delay differential equations-based). This model shows an endemic and unstable equilibrium point, because there is always a prevalescence of the disease in the population. More complicated compartmental structure is possible. For example, there are
$SEIR$ and $SEIS$ models, with an exposed period between being infected and becoming infective \cite{bib:brauer2013}. However, it makes diff\/icult the assignment of resources for treatments \cite{bib:brauer2013}.

\par The $SIRS$ model is given by the following system of delay differential equations:
\begin{equation}
\begin{cases} \frac{dS}{dt}=-\alpha S(t) I(t) + \beta I(t-\omega),  \\
\frac{dI}{dt}=\alpha S(t) I(t) - \beta I(t), \label{SIRSeq} \\
\frac{dR}{dt}=\beta I(t) - \beta I(t-\omega),
\end{cases}
\end{equation}

\par where $\alpha \mbox{, }\beta \mbox{ y }\omega$ are called \textit{infection rate}, \textit{recovery rate} and \textit{temporary immunity period length}, respectively.

As it is seen in epidemics spreading, rumors can also exhibit different behaviors, such as a behavior where everybody is talking about the same topic over and over, but after some time, this attitude ends and people lose their interest about this topic very quickly.  So, not everybody is the same and there will be different transmission rates of the rumor among different people (i.e., if there was heterogeneity in classes). Hence a $SIR$-type model can be used to study these kind of situations (see, for example, \cite{bib:thompsonestrada}). Rumors with a periodic spreading behavior also exist, these can be events that are expected to happen with a certain frequency. In this paper to model periodicity the $SIRS$ with constant temporary immunity model is used.

\subsection{Adaptative Neuro-Fuzzy Inference Systems (ANFIS)}

Fuzzy logic has been used as a powerful tool for modeling uncertainty \cite{bib:zimmermann}. 
Here we will consider using the well-known fuzzy Takagi-Sugeno model, which arises as an effort for systematically generating linguistic rules from an input-output data \cite{bib:fuzzymatlab}.

In order to provide a good f\/itting to a dataset, a special neural network called ANFIS is used. It is functionally equivalent to a radial basis function network (RBFN) under certain conditions \cite{bib:neurofuzzy}, but unlike the RBFN, ANFIS allows to do qualitative research of the data because of its close relations to fuzzy logic. ANFIS was proposed to be equivalent to the Takagi-Sugeno model \cite{bib:neurofuzzy}, so that it is possible to estimate the linguistic rules of the Takagi-Sugeno model starting from the training of an ANFIS network.



For the simplest of explanations of the layers, a two-input linear Takagi-Sugeno model is considered (see 
  \cite{bib:neurofuzzy} for details):

\begin{itemize}
	\item \textbf{Layer 1 (input layer)}: a membership function is applied to the inputs. This will be the output function of the f\/irst layer:
	
	\begin{equation}	\label{layer1}
	O_{1,i}=\begin{cases} \mu_{\tilde{A}_i}(x) & \mbox{if } i=1,2,\\
   \mu_{\tilde{B}_{i-2}}(y) & \mbox{if } i=3,4,\end{cases}
	\end{equation}
	
	\par where $x$ (or $y$) are the input elements of the node $i$ and $\tilde{A}_i$ (or $\tilde{B}_{i-2}$) are membership values between 0 and 1. $O_{1,i}$ is the equivalent of a fuzzif\/ication. It can be considered as a membership function (or MF for short), for example, a generalized bell-shaped MF:
	
	\begin{equation} \label{layer1eq}
	\mu_{\tilde{A}}=\frac{1}{1+{|\frac{x-c_i}{a_i}|^{2b_i}}},
	\end{equation}
	\par where $\{a_i,b_i,c_i\}$ are the function's parameters known as \textbf{premise parameters}.
	
	\item \textbf{Layer 2}: every node, denoted by $\Pi$, has an output function: \\
	\begin{equation}  \label{layer2}
	O_{2,i}=w_i=\mu_{\tilde{A}_i}(x) \mu_{\tilde{B}_i}(y), \mbox{ with } i=1,2.
	\end{equation}
	
	\par This output represents the f\/iring strength of a rule; that is to say, the degree on which a rule is satisf\/ied. Whenever it is closest to 1, it will be more fulf\/illed, however, if this value is close to 0, there will be low fulf\/ilment.
	
	
	\item \textbf{Layer 3}: every node, denoted by $N$, has as output function the \textbf{normalized f\/iring strengths} (degree of fulf\/illment of a fuzzy rule with maximum value 1)
	
	\begin{equation} \label{layer3}
	O_{3,i}=\bar{w}_i=\frac{w_i}{w_1+w_2} \mbox{, with } i=1,2.
	\end{equation}
	
	\item \textbf{Layer 4}: the outputs of the fuzzy rules are weighted by the normalized f\/iring strenghts, so the output function is given by
	
	\begin{equation} \label{layer4}
	O_{4,i}=\bar{w}_i f_i=\bar{w_i}(p_ix + q_iy + r_i),
	\end{equation}
	
	\par where $\bar{w_i}$ is the normalized f\/iring strength calculated in the third layer, and $\{p_i,q_i,r_i\}$ are the parameters of this node known as  \textbf{consequent parameters}.
	
	\item \textbf{Layer 5 (output layer)}: the only node of this layer, denoted by $\Sigma$, computes the general output of the network:
	\begin{equation} \label{layer5}
	O_{5,1}=\sum_{i}\bar{w_i}f_i=\frac{\sum_{i}w_if_i}{\sum_{i}w_i}.
	\end{equation}
	
\end{itemize}

\par This is not the unique setting of ANFIS, because it is possible to merge the third and fourth layers in order to obtain a four-layer network \cite{bib:neurofuzzy}. From the ANFIS architecture we note that when the values of the premise parameters are fixed, the overall
output can be expressed as a linear combination of the consequent parameters.
Moreover, the weight normalization can be done in the output layer \cite{bib:neurofuzzy}. This network can be trained using a hybrid learning algorithm (see \cite[\S 8.5]{bib:neurofuzzy}): the output layer parameters are estimated using least mean squares and the rest of the network is trained using the backpropagation method \cite{bib:neurofuzzy}. Indeed, in the forward pass of the hybrid learning algorithm, node outputs go forward until layer~4 and the consequent parameters are identified by the least-square method. In the backward pass, the error signals propagate backward and the premise parameters are updated by gradient descent. The consequent parameters thus identified are optimal under the condition that the premise parameters are fixed \cite[\S 8.5]{bib:neurofuzzy}).

For this reason, the hybrid approach converges much faster since it reduces the search space dimensions of the original pure backpropagation
 method. Thus we should always look for the possibility of decomposing the parameter set in the first place. For ANFIS, this can be achieved
  if the MF on the consequent part of each rule is replaced by a piecewise linear approximation with two consequent parameters \cite[\S 12.3]{bib:neurofuzzy}).

%% file: Segmen.tex
\section{Fuzzy logic approach using deterministic models}

Deterministic models have been frequently used to analyze the qualitative behavior of phenomena. In general, these kinds of models are much easier to analize than stochastic models due to their definite nature \cite{bib:abraham2009computational}. Most of these models are restricted to the assumption of constant rates \cite{bib:roberts2015}, therefore they often fail to model important shifts that are seen in real situations (see for example, the experiments in Sections 4.1 and 4.2). In this paper we propose the ANFIS approach using deterministic models, in order to provide a good approximation of the general behavior of the spreaders, while preserving the qualitative information that can be extracted from the deterministic models.

The strategy follows two important steps:

\begin{enumerate}
	\item \textbf{Fitting data using a deterministic model:} a deterministic model is picked so that the error between real data and estimated data is as small as possible. This model will estimate unknown variables that should be implicit to the dataset.
	\item \textbf{Training the ANFIS network:} this is carried out in two stages: first, the input variables should be chosen based on a criterion (e.g., Mallows' $C_p$). Then, the real dataset must be split into train and test data. When training ANFIS overf\/itting (or overlearning, which is frequently encountered in complicated adaptive systems) must be avoided by selecting right input variables, amount of MFs, and number of epochs.
\end{enumerate}

In this paper we considered two rumor-related datasets, which we model using some well-known deterministic models. The first one, \textquotedblleft Gangnam Style'', shows a similar behavior to the $SIR$ spreaders curve. The second rumor, \textquotedblleft Game of Thrones'', has a periodical behavior and therefore, we considered a $SIRS$ model with delay. If a dataset is non-chaotic, it is possible to grasp the fundamental behavior of the trending topic, and then supply an adequate deterministic model for the situation.

This approach allows us to analyze the data via deterministic models and, at the same time, by using ANFIS.

\section{Numerical Applications}

We accessed Google Trends, the public web facility of Google Inc. based on Google Search engine in order to obtain weekly measured trending topics. Two trending topics were selected: \textquotedblleft Gangnam Style" and \textquotedblleft Game of Thrones". 
The reason to use Google Trends is that the Google search engine is one of the major sources of information worldwide, 
 regardless of the veracity of what is posted on it.
The simulations and numerical results were provided by the scientif\/ic computation tool known as MATLAB, version R2013a. Fuzzy Logic Toolbox within the framework of MATLAB was used for the training and evaluation of ANFIS. The functions used were the following: \texttt{genfis1} (Generate Fuzzy Inference System structure from data using grid partition) for generating a Sugeno-type FIS structure (used as initial conditions), which is later trained using \texttt{anfis}, and finally \texttt{evalfis}, for evaluating the experimental data set, \texttt{data}, on our trained system \cite{bib:Matlabh}.

Populations of ignorants, spreaders and stifflers were normalized to a [0-1] scale. It is considered that spreaders are individuals who browse the internet looking for information related to the topic they are interested in.

In both experiments, each dataset was split into 50:50 for training and testing data, respectively. The division of the datasets can be seen in Figure \ref{fig:gangnamanfis}(c) and Figure \ref{fig:gotfis}(c). The authors could also have chosen for the training and the tests any of the options 60:40 or 70:30; however, we seek that our numerical experiments have, as far as possible, a good visualization.

\subsection{The \textquotedblleft Gangnam Style" fever}

\begin{figure}[H]
	\centering
	\includegraphics[width=1\linewidth]{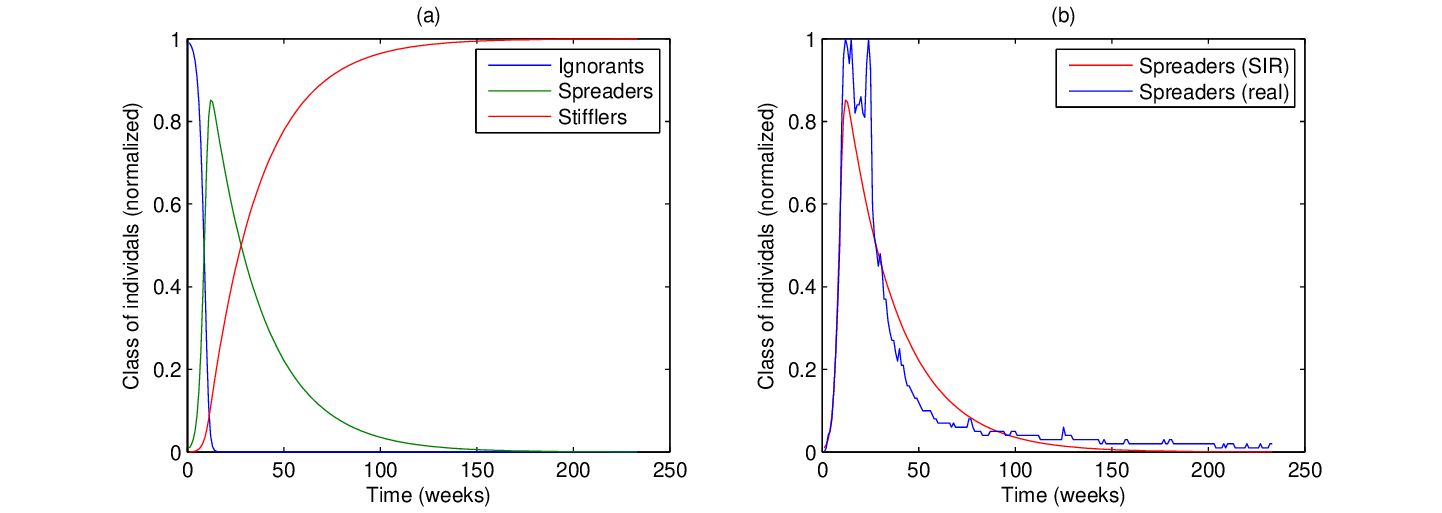}
	\caption{Class of individuals related to the trending topic \textquotedblleft Gangnam Style''. (a) $SIR$ model adapted to \textquotedblleft Gangnam Style''; (b) graphical comparison between real and estimated data by the $SIR$ model.}
	\label{fig:GangnamStyleSIR}
\end{figure}

\textquotedblleft Gangnam Style" is a song interpreted by the South Korean musician Psy. It was released on July 15, 2012 together with a music video, which shortly became the most viewed video of Youtube. After several weeks, this popular song was not interesting anymore for the majority of the web users. As it is seen in Figure \ref{fig:GangnamStyleSIR}(b), there are two important peaks around the first 50 weeks, this is a behavior known as \textquotedblleft chimney" \cite{bib:kairam}.\textquotedblleft Chimneys" are composed by two \textquotedblleft wedges", (where population rises and falls at similar rates over time) which are separated by a level in which the population does not have a strong variation. In other words, there is a rumor outbreak but the user interest remains very high for a few weeks before dropping down. It is suggested in \cite{bib:kairam} that the majority of topics share the same patterns in both search and social media activity, so modeling this situations using a model of social behavior model would be appropriate. The $SIR$ deterministic model only has one peak, therefore it is used for studying only the general behavior of this rumor and extracting useful information for predictions.

Google Trends dataset provided dates against search percents, so the only available data consisted of time and spreaders variables. A better dataset was needed to train ANFIS properly, so by applying an $SIR$ model to the spreader data would create estimated curves for ignorants, stifflers and also spreaders (all time-dependent). In order to f\/it the $SIR$ model, we needed to solve the parameters searching problem, which was solved using exhaustive search on the $\alpha$ and $\beta$ parameters from 0.001 to 1, taking a 0.001 length step. For each value of $\alpha$ and $\beta$, Euler's method was used to solve the ODE with initial conditions $S_0=0.99,I_0=0.01,R_0=0$. And so, the best $SIR$ model was chosen to minimize the $LSE$ (Least Squares Error) of the spreaders:

\begin{equation*}
LSE=\frac{1}{n}\sum_{i=1}^{n}(\hat{Y_i}-Y_i)^2
\end{equation*}

where $\hat{Y_i}$ is the estimated value and $Y_i$ is the real value. The elapsed time of this algorithm was 0.958143 seconds, for $n=233$. The parameters are shown in Table \ref{table:parametrosSIRGS}. Figure \ref{fig:GangnamStyleSIR} shows the closest f\/it provided by an $SIR$ model to this dataset.

\begin{table}[H]
	\centering
	\begin{tabular}{l*{6}{c}r}
		             & $LSE$ & $I_0$ & $\alpha$ & $\beta$ & $n$ \\
		\hline
		Value & 0.0046 & 0.01 & 0.7750 & 0.0380  & 233 \\
	\end{tabular}
	\caption{\textquotedblleft Gangnam Style'' $SIR$ parameters.}
	\label{table:parametrosSIRGS}
\end{table}

Now we have three variables to select as an input: time, ignorants ($SIR$), and stifflers ($SIR$), but it is necessary to select the appropriate variables that can allow ANFIS to produce both a well trained model with good accuracy. For this task we considered Mallows' $C_p$ criterion, from which the $C_p$ statistic is defined as follows: 

\begin{equation*}
C_p=\frac{\sum_{i=1}^{n}(\hat{Y_i}-Y_i)^2}{\sigma_d^2}-n+2p,
\end{equation*}

\noindent where $\sigma_d^2$ is the mean square error of the residuals for a full model using all possible input variables, $n$ is the size of the dataset, and $p$ is the number of parameters of the model \cite{bib:mayinputs}. ANFIS has two sets of parameters: nonlinear and linear, which are determined by the number of input variables, membership function tipe and amount, and the network's output, respectively. So, $p$ can be calculated using the following equation:

\begin{equation*}
p=\begin{cases} VK\theta + K^V & \mbox{for constant output}, \\
VK\theta + (V+1)K^V & \mbox{for linear output},\end{cases}
\end{equation*}

where $V$ is the number of input variables, $K$ is the number of membership functions, and $\theta$ is the number of parameters of the membership function. The first and the second terms correspond to the number of nonlinear and linear parameters, respectively. Theoretically, the value of $C_p$ for the optimal model will be $p$ \cite{bib:mayinputs}, therefore, in order to select the best model, we first calculated the Mallows' $C_p$. Then we computed the distance between this result and each $p$ in order to find which model's coefficient is the closest to its number of parameters.

Table \ref{tab:corrGangnamStyle} suggests that time and stifflers ($SIR$) with constant output provide the optimum model. Hence, we selected these two variables as the ANFIS inputs.

\begin{table}[H]
	\centering
\begin{tabular}{cccc}
	\hline
	\multicolumn{1}{|c|}{Output}   & \multicolumn{3}{c|}{Input variables (single-input)}                                                                                                                                                                                                                              \\ \hline
	\multicolumn{1}{|c|}{}         & \multicolumn{1}{c|}{ignorants ($SIR$)}                                                             & \multicolumn{1}{c|}{stifflers ($SIR$)}                                                  & \multicolumn{1}{c|}{time}                                                             \\ \hline
	\multicolumn{1}{|c|}{constant} & \multicolumn{1}{c|}{\num{1.1381e4}}                                                           & \multicolumn{1}{c|}{118.5243}                                                         & \multicolumn{1}{c|}{\num{7.0423e3}}                                                 \\ \hline
	\multicolumn{1}{|c|}{linear}   & \multicolumn{1}{c|}{\num{5.5173e5}}                                                            & \multicolumn{1}{c|}{778.0769}                                                         & \multicolumn{1}{c|}{\num{3.6166e4}}                                                 \\ \hline
	&                                                                                                  &                                                                                       &                                                                                       \\ \hline
	\multicolumn{1}{|c|}{Output}   & \multicolumn{3}{c|}{Input variables (double-input)}                                                                                                                                                                                                                              \\ \hline
	\multicolumn{1}{|c|}{}         & \multicolumn{1}{c|}{\begin{tabular}[c]{@{}c@{}}ignorants ($SIR$) +\\ stifflers ($SIR$)\end{tabular}} & \multicolumn{1}{c|}{\begin{tabular}[c]{@{}c@{}}ignorants ($SIR$) +\\ time\end{tabular}} & \multicolumn{1}{c|}{\begin{tabular}[c]{@{}c@{}}stifflers ($SIR$) +\\ time\end{tabular}} \\ \hline
	\multicolumn{1}{|c|}{constant} & \multicolumn{1}{c|}{65.62}                                                                       & \multicolumn{1}{c|}{\num{3.9308e3}}                                                 & \multicolumn{1}{c|}{\num{48.4960}}                                                  \\ \hline
	\multicolumn{1}{|c|}{linear}   & \multicolumn{1}{c|}{\num{1.3748e3}}                                                            & \multicolumn{1}{c|}{\num{3.2994e3}}                                                 & \multicolumn{1}{c|}{520.2115}                                                         \\ \hline
\end{tabular}
	\caption{Distance between Mallows' $Cp$ and the corresponding number of parameters of each model. This table was considered for choosing the best ANFIS input variables for the \textquotedblleft Gangnam Style'' experiment.}
	\label{tab:corrGangnamStyle}
\end{table}

\begin{figure}[H]
	\centering
	\includegraphics[width=1\linewidth]{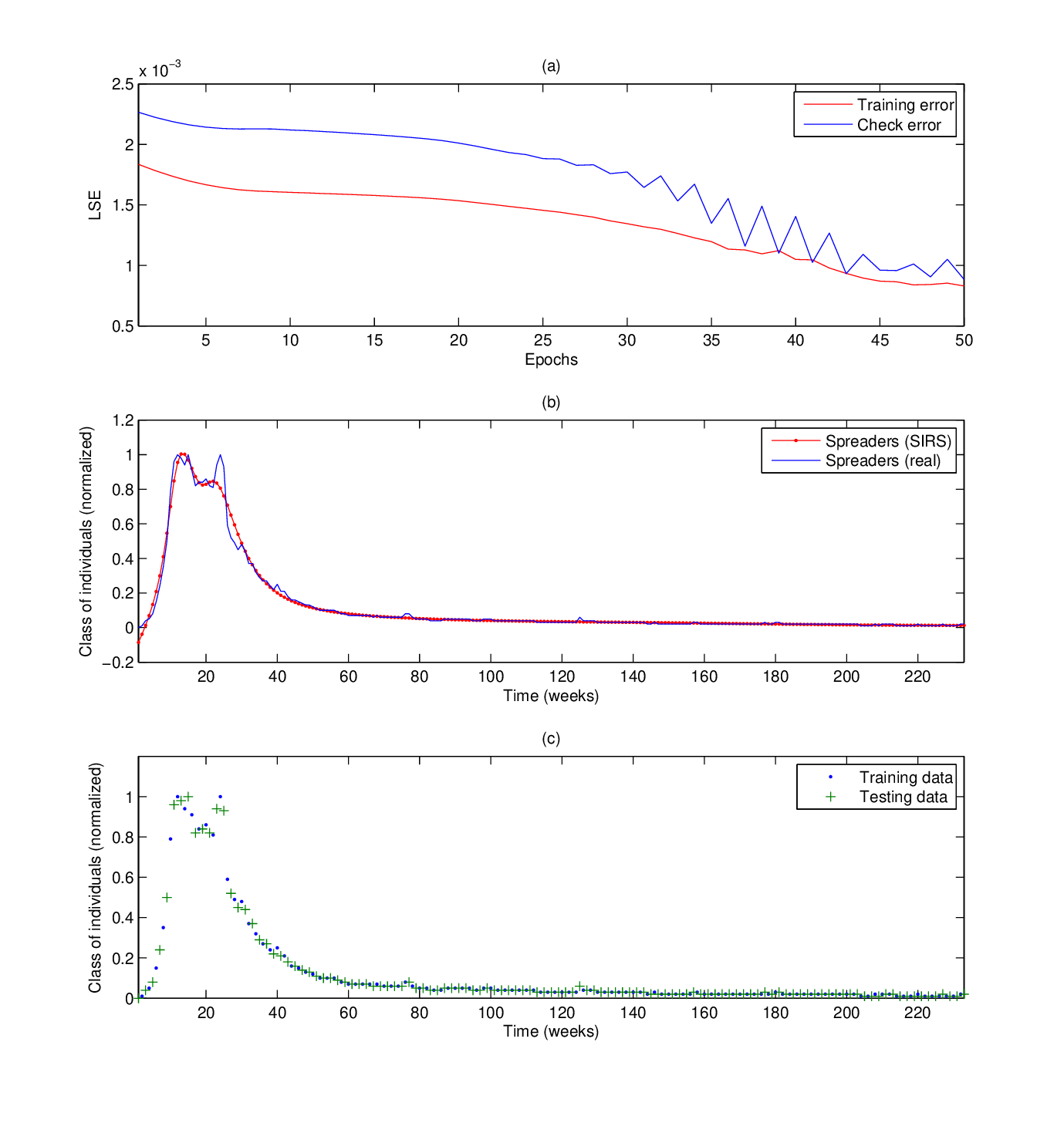}
	\caption{ANFIS \textquotedblleft Gangnam Style" prediction curve analysis. Input variables: time and stifflers ($SIR$), constant output. (a) Comparison of training and check $LSE$ for constant output, (b) comparison of real data and ANFIS output, (c) division of the dataset into training data and testing data.}
	\label{fig:gangnamanfis}
\end{figure}

As a neural network, ANFIS also requires to be trained within a certain number of epochs. Different training epochs can affect the capability of the network to classify patterns on which they have not been trained (generalization capability); for example, an overtrained network will only produce good results with familiar data. The real data was split into a training data and a validation data. Several simulations were done using different number of training epochs for ANFIS, verifying that the check error (difference between ANFIS output and validation data) and training errors (difference between ANFIS output and training data) were close (so that ANFIS would have generalization capability). Also, we compared the the generalization capabilities and $LSE$ between linear and constant output models. Finally, the best ANFIS network for this dataset was found within 50 epochs, as it can be seen in Figure \ref{fig:gangnamanfis}.

Membership functions also play a fundamental role in Fuzzy Inference Systems modelling, and it is important that these are chosen carefully. The performance of ANFIS is very sensitive to the amount of MFs in the system, so it should be expected that complex systems with a great quantity of MFs will perform poorly because of the amount of the premise parameters that need to be estimated.
MFs can have many shapes, such as triangle, trapezoidal, gaussian, generalized bell-shaped, etc. \cite{bib:neurofuzzy}. The shape of the MF can affect the accuracy of the task that is given to the ANFIS, and according to the practical results of \cite{bib:mfcomparepred} and \cite{bib:mfcompareclass}, the Gassian MF provided a higher degree of accuracy with less complexity (this MF only has two parameters). The Gaussian MF is defined as follows:

\begin{equation*}
\mu_{\tilde{A}}=exp(-\frac{(x-m)^2}{2\sigma^2}),
\end{equation*}
\par where $m$ and $\sigma$ represent the center and the width of the MF, respectively.

Regarding the fuzzy rules generated by ANFIS, three Gaussian MFs were selected per input variable, so that increasing the inputs to our network would not cause the number of parameters to quickly surpass the available data. The fuzzy rules are used only for computational purposes (see \cite{bib:fuzzymatlab} for basic details on fuzzy inference).

Although the spreaders curve provided by the $SIR$ model fits the most general behavior of the data (Figure \ref{fig:GangnamStyleSIR}), the ANFIS output can now fit the important shift that happens between the $20^{th}$ and $30^{th}$ week, as it is seen in Figure \ref{fig:gangnamanfis}.

\subsection{Seasons of \textquotedblleft Game of Thrones"}

\begin{figure}[h]
	\centering
	\includegraphics[width=1\linewidth]{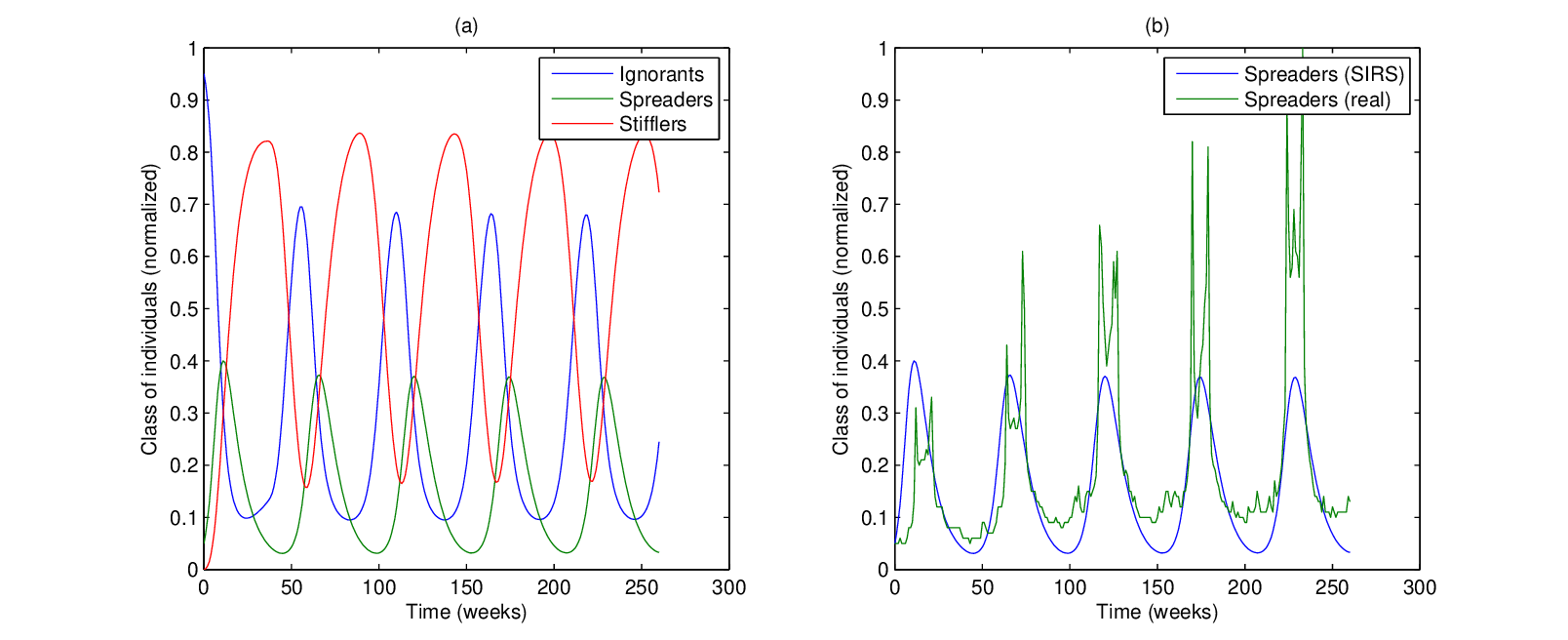}
	\caption{Class of individuals related to \textquotedblleft Game of Thrones\textquotedblright. (a) Delayed $SIRS$ populations; (b) delayed $SIRS$ spreaders compared to real spreaders of \textquotedblleft Game of Thrones''.}
	\label{fig:sirsgot}
\end{figure}

Game of Thrones is a critically acclaimed television series, which has been highly anticipated throughout the years because of its great success. The Google Trends dataset shows a periodical \textquotedblleft chimney'' which happens between each season premiere and finale (see Figure \ref{fig:sirsgot}(b)). Another interesting feature of its success is the increasing number of viewers throughout the years.

In order to obtain weekly data, Seasons 2 to 6 were considered for this study. Contrary to the \textquotedblleft Gangnam Style" trend, the interest in this television series is not lost, instead, every year numerous people start watching this show every time a new season is premiered.

We used SIRS model with delay to f\/it the spreaders again, but this time f\/itting the parameters with exhaustive search was not convenient because of the broad amount of values that $\omega$ (i.e., the temporary immunity period length) may take. Instead, we used the MATLAB function \texttt{fminsearch}, which is based on the algorithm of Nelder-Mead described in \cite{bib:fminsearch}. We proceed to minimize the function:

\begin{equation*}
Err(\alpha,\beta,\omega)=\|Y(t)-\hat{Y}(\alpha,\beta,\omega,t)\|_2^2=
\sum_{t=0}^{n-1}(Y(t)-\hat{Y}(\alpha,\beta,\omega,t))^2,
\end{equation*}

\par where $Y$ is the real data, $\hat{Y}$ are the estimated spreaders by the delayed $SIRS$ model, and $n$ is the number of measures, such that we have a sequence of $n$ elements: $t=0,1,2,...,n-2,n-1$. For each iteration of Nelder-Mead's method, DDE's are solved using the MATLAB function \texttt{ddesolve} \cite{bib:dde23}.

Since the Nelder-Mead algorithm is a direct method, it has the important advantage that no information about the gradient is needed \cite{bib:nongradstandord}. \texttt{fminsearch} does not allow to f\/ind global minima for any initial point, thus, we had to start suff\/iciently close to the global minimum. Finding the global minimum is equivalent to finding the best fitting SIRS model for the real data. After manually changing the initial parameters, we found that for $\alpha_0=0.7$, $\beta_0=0.5$, $\omega=40$, we obtained the optimum parameters, shown in Table \ref{table:parametrosSIRSgot} in 9.527718 seconds.

In Figure \ref{fig:sirsgot} the f\/itting quality is compared to the real data. It can be seen in this figure that the spreaders provided by the $SIRS$ have constant amplitude, however, the amplitude of the real data oscilations increases through time. Therefore, the f\/itting quality was not the best.

\begin{table}[H]
	\centering
	\begin{tabular}{l*{6}{c}r}
		              & LSE & $I_0$ & $\alpha$ & $\beta$ & $\omega$ & $n$\\
		\hline
		Value & 0.0153 & 0.05 & 0.4981 & 0.1475 & 37.1562 & 261\\
	\end{tabular}
	\caption{Delayed $SIRS$ parameters for \textquotedblleft Game of Thrones\textquotedblright.}
	\label{table:parametrosSIRSgot}
\end{table}

To provide a good general f\/itting for this complex curve, the estimated number of spreaders was considered as a possible input, as it is shown in Table \ref{tab:corrgot}, which displays the distance between Mallows' $C_p$ and the number of parameters.

\begin{table}[H]
	\centering
\begin{tabular}{cccclll}
	\cline{1-5}
	\multicolumn{1}{|c|}{Output}   & \multicolumn{4}{c|}{Input variables (single-input)}                                                                                                                                                                                                                                                                                                                                                                                                               &  &  \\ \cline{1-5}
	\multicolumn{1}{|c|}{}         & \multicolumn{1}{c|}{ignorants ($SIRS$)}                                                                                   & \multicolumn{1}{c|}{spreaders ($SIRS$)}                                                                       & \multicolumn{1}{c|}{stifflers ($SIRS$)}                                                                       & \multicolumn{1}{c|}{time}                                                                                   &  &  \\ \cline{1-5}
	\multicolumn{1}{|c|}{constant} & \multicolumn{1}{c|}{\num{1.1512e3}}                                                                                           & \multicolumn{1}{c|}{472.2731}                                                                                & \multicolumn{1}{c|}{187.4789}                                                                                & \multicolumn{1}{c|}{\num{1.1637e3}}                                                                               &  &  \\ \cline{1-5}
	\multicolumn{1}{|c|}{linear}   & \multicolumn{1}{c|}{175.5956}                                                                                            & \multicolumn{1}{c|}{213.1175}                                                                                & \multicolumn{1}{c|}{185.5772}                                                                                & \multicolumn{1}{c|}{170.4134}                                                                                &  &  \\ \cline{1-5}
	&                                                                                                                         &                                                                                                             &                                                                                                             &                                                                                                             &  &  \\ \cline{1-4}
	\multicolumn{1}{|c|}{Output}   & \multicolumn{3}{c|}{Input variables (double-input)}                                                                                                                                                                                                                                                                                                 &                                                                                                             &  &  \\ \cline{1-4}
	\multicolumn{1}{|c|}{}         & \multicolumn{1}{c|}{\begin{tabular}[c]{@{}c@{}}ignorants ($SIRS$) +\\ stifflers ($SIRS$)\end{tabular}}                      & \multicolumn{1}{c|}{\begin{tabular}[c]{@{}c@{}}ignorants ($SIRS$) +\\ spreaders ($SIRS$)\end{tabular}}          & \multicolumn{1}{c|}{\begin{tabular}[c]{@{}c@{}}spreaders ($SIRS$) +\\ stifflers ($SIRS$)\end{tabular}}          &                                                                                                             &  &  \\ \cline{1-4}
	\multicolumn{1}{|c|}{constant} & \multicolumn{1}{c|}{268.7067}                                                                                            & \multicolumn{1}{c|}{437.7463}                                                                                & \multicolumn{1}{c|}{315.8023}                                                                                &                                                                                                             &  &  \\ \cline{1-4}
	\multicolumn{1}{|c|}{linear}   & \multicolumn{1}{c|}{196.2437}                                                                                             & \multicolumn{1}{c|}{190.5610}                                                                                 & \multicolumn{1}{c|}{198.7697}                                                                                 &                                                                                                             &  &  \\ \cline{1-4}
	\multicolumn{1}{|c|}{}         & \multicolumn{1}{c|}{\begin{tabular}[c]{@{}c@{}}time +\\ ignorants ($SIRS$)\end{tabular}}                                  & \multicolumn{1}{c|}{\begin{tabular}[c]{@{}c@{}}time +\\ spreaders ($SIRS$)\end{tabular}}                      & \multicolumn{1}{c|}{\begin{tabular}[c]{@{}c@{}}time +\\ stifflers ($SIRS$)\end{tabular}}                      &                                                                                                             &  &  \\ \cline{1-4}
	\multicolumn{1}{|c|}{constant} & \multicolumn{1}{c|}{972.5523}                                                                                            & \multicolumn{1}{c|}{145.5336}                                                                                 & \multicolumn{1}{c|}{721.7847}                                                                                &                                                                                                             &  &  \\ \cline{1-4}
	\multicolumn{1}{|c|}{linear}   & \multicolumn{1}{c|}{165.4337}                                                                                             & \multicolumn{1}{c|}{204.7196}                                                                                 & \multicolumn{1}{c|}{179.5269}                                                                                 &                                                                                                             &  &  \\ \cline{1-4}
	\multicolumn{1}{l}{}           & \multicolumn{1}{l}{}                                                                                                    & \multicolumn{1}{l}{}                                                                                        & \multicolumn{1}{l}{}                                                                                        &                                                                                                             &  &  \\ \cline{1-5}
	\multicolumn{1}{|c|}{Output}   & \multicolumn{4}{c|}{Input variables (triple-input)}                                                                                                                                                                                                                                                                                                                                                                                                               &  &  \\ \cline{1-5}
	\multicolumn{1}{|c|}{}         & \multicolumn{1}{c|}{\begin{tabular}[c]{@{}c@{}}ignorants ($SIRS$) +\\ spreaders ($SIRS$) +\\ stifflers ($SIRS$)\end{tabular}} & \multicolumn{1}{c|}{\begin{tabular}[c]{@{}c@{}}time +\\ ignorants ($SIRS$) +\\ spreaders ($SIRS$)\end{tabular}} & \multicolumn{1}{c|}{\begin{tabular}[c]{@{}c@{}}time +\\ ignorants ($SIRS$) +\\ stifflers ($SIRS$)\end{tabular}} & \multicolumn{1}{c|}{\begin{tabular}[c]{@{}c@{}}time +\\ spreaders ($SIRS$) +\\ stifflers ($SIRS$)\end{tabular}} &  &  \\ \cline{1-5}
	\multicolumn{1}{|c|}{constant} & \multicolumn{1}{c|}{298.0590}                                                                                             & \multicolumn{1}{c|}{61.0377}                                                                                 & \multicolumn{1}{c|}{142.8159}                                                                                 & \multicolumn{1}{c|}{78.1933}                                                                                 &  &  \\ \cline{1-5}
	\multicolumn{1}{|c|}{linear}   & \multicolumn{1}{c|}{96.6029}                                                                                             & \multicolumn{1}{c|}{123.2538}                                                                                 & \multicolumn{1}{c|}{124.3661}                                                                                 & \multicolumn{1}{c|}{123.7143}                                                                                 &  &  \\ \cline{1-5}
\end{tabular}
	\caption{Distance between Mallows' $Cp$ and the corresponding number of parameters of each model. This table was considered for choosing the best ANFIS input variables for the \textquotedblleft Game of Thrones\textquotedblright experiment.}
	\label{tab:corrgot}
\end{table}

\begin{figure}[H]
	\centering
	\includegraphics[width=1\linewidth]{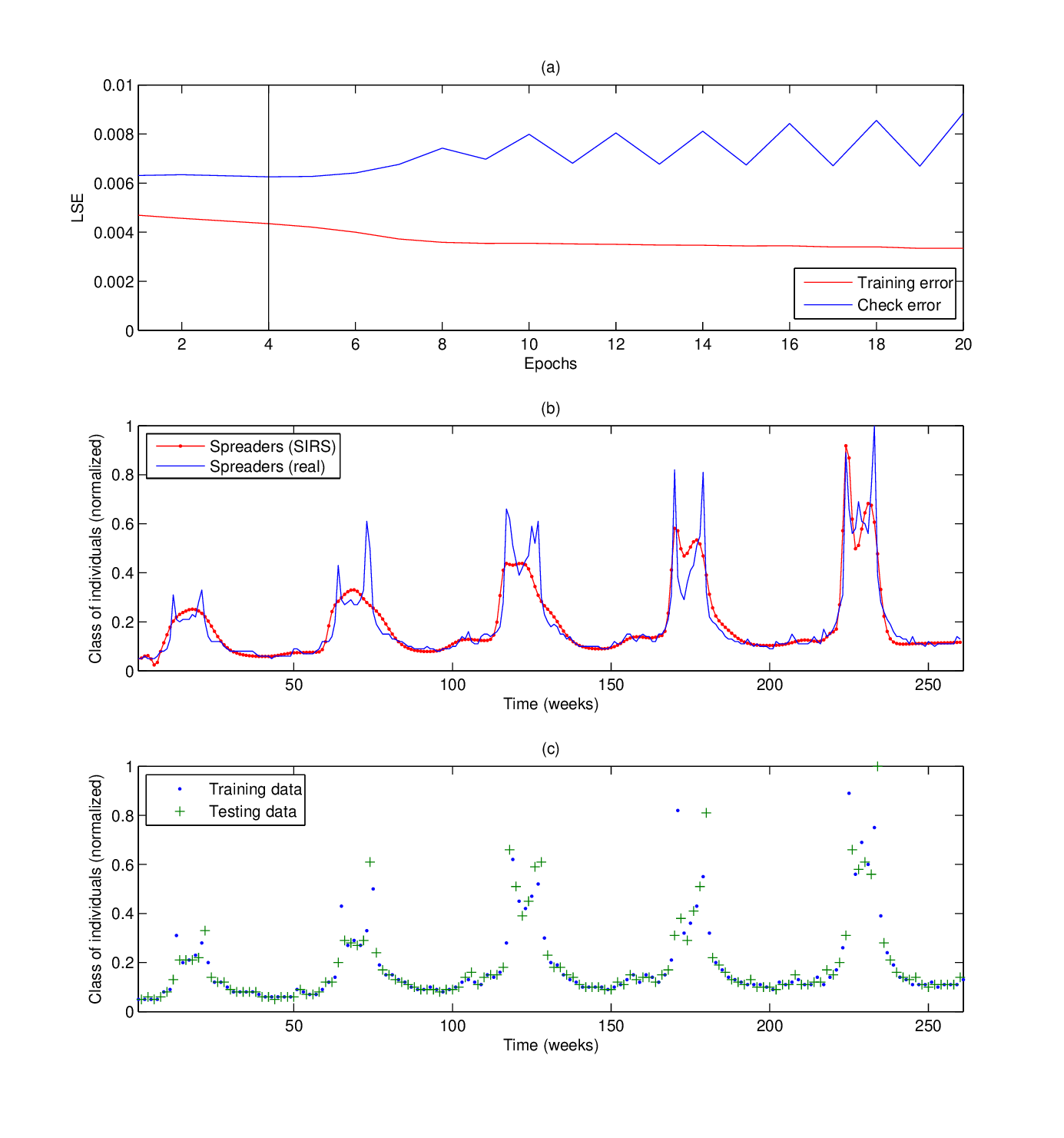}
	\caption{ANFIS \textquotedblleft Game of Thrones" prediction curve analysis. Input variables: time, ignorants (SIRS) and spreaders (SIRS), constant output. (a) Comparison of training and check $LSE$ for constant output, (b) comparison of real data and ANFIS output, (c) division of the dataset into training data and testing data.}
	\label{fig:gotfis}
\end{figure}

Time, spreaders ($SIRS$), and stifflers (SIRS) were used as input variables for ANFIS. After many simulations, the best model was obtained with the outstandingly low amount of four epochs, with three Gaussian MFs (just like the \textquotedblleft Gangnam Style'' example), and constant output (see Figure \ref{fig:gotfis}). In this experiment, the overfitting problem appears, this is because ANFIS was trained with few data points compared to its great number of parameters \cite{bib:anderson}, which will increase depending on the amount of input variables and MFs (see \cite{bib:chiang1995} for details). As it can be seen in Figure \ref{fig:gotfis}(a), the Epochs were limited to only four, because otherwise, the overfitting would have been greater. It is expected to be an evident difference between training and check error because of the lack of data found on the peaks (see Figure \ref{fig:gotfis}(c)). We believe that a more accurate deterministic model that fits the spreaders dataset can be found. If so, the relation between the simulations and the real data will make it easier for a much simpler network to avoid overfitting while providing an accurate model.

This result shows that f\/itting models for even more complex behaviors such as periodicity with nonconstant amplitude can be found, using the procedure proposed in this paper. In this case, a delayed differential equations-based model provided a good estimation of unknown variables, and with the help of fuzzy rules (generated by ANFIS) we could f\/ind an appropriate general curve for the spreaders. 

%% file: Conclus.tex
\section{Concluding remarks}

In the course of time, different techniques have been used to study the spreading of rumors. Fitting a dataset is a problem to which many approaches have been proposed \cite{bib:hornikMLN}; in particular, the discussed use of ANFIS in this paper, broadens up the f\/ield for new investigations.

The ANFIS approach allowed us to f\/ind the best curve that explains the general behavior of each dataset, however, this network is very sensible to the number of parameters, so we selected an appropriate type and amount of MFs. Mallows $C_p$ criterion can be useful for selecting optimum prediction models.

After performing many tests and analyzing the process of f\/itting the trending datasets, we concluded that if a dataset does not provide enough information, it is possible to obtain implicit information by applying known mathematical models (particularly, deterministic models). Knowing these variables allowed us to f\/it the dataset by training the ANFIS network. This has been one of the most useful results of this work and will be the starting point for future research.